\documentclass[aps,twocolumn,showkeys]{revtex4}
\usepackage{dcolumn}
\usepackage{bm}
\usepackage{amssymb}
\usepackage{amsfonts}
\usepackage{amsmath}
\usepackage{graphicx}
\begin{document}

\title[LTEE]{Modeling evolution in a Long Time Evolution Experiment with E. Coli}
\author{Dario A. Le\'{o}n}
\affiliation{ICIMAF}
\author{Augusto Gonz\'{a}lez}
\affiliation{ICIMAF}


\begin{abstract}
Taking into account an evolutionary model of mutations in term of Levy flights that was previously constructed, we designed an algorithm to reproduce the evolutionary dynamics of the Long-Term Evolution Experiment (LTEE) with E. Coli bacteria. The algorithm enables us to simulate mutations under natural selection conditions. The results of simulations on competition of clones, mean fitness, etc., are compared with experimental data. We attained to reproduce the behavior of the mean fitness of the bacteria cultures, get our own interpretations and more tuned descriptions of some phenomena taking part within the experiment, such as fixation and drift processes, clonal interference and epistasis.
\end{abstract}

\maketitle

\section{Introduction}

Mutations, as the ultimate source of heritable variation, joined with
natural selection leads to biological evolution. In the Long-Term Evolution
Experiment (LTEE) with E. Coli cultures \cite{Sitio_Lenski}, as in any typical evolution
process, the individuals of a given population (bacteria in this case)
compete one with the others to survive in a controlled environment. The
process could be well characterized by the fitness parameter of the
phenotype of each individual, which consist in the product of two factors
account for its capacity to survive and reproducing trough the next
generation \cite{Notes_Fitness}. Regarding fitness, mutations just can be three kinds: beneficial
(high fitness), neutral (same fitness) or deleterious (decreased fitness).
Then, the phenotypes with highest fitness will be picked by natural selection
and will have more representability in new generations. Of course, the
dynamic of the bacteria cultures depends on the distribution of fitness
effects (DFE) of the arising mutations, they define the range of possible
evolutionary trajectories a population can follow \cite{DFE}. %If we control the
%population size, as in the experiment, evolution can be viewed as an
%optimization problem of the fitness \cite{DarioyAugusto}. In this picture
Obviously not all mutations
survive next generation, those that do not are called driver mutations
and the others passengers. Some of the passengers with high benefit can be
reproduced until they are present in most members of the culture after a
certain number of generations, these mutations are said to be fixed in the
population \cite{Adaptation_Lenski} and because of that are of utmost importance.

If we look at the DNA molecule, mutations can be as simple as little
mistakes in the replication process surviving the repair mechanisms. This
base replacements are called point mutations. But mutations also can imply
radical rearrangements in the DNA strand, which have a significant different
behavior \cite{DarioyAugusto}. In a previously work \cite{DarioyAugusto} we propose a
Levy model for the accumulation of mutation along a cell lineage in which
the rate of the second event is one-third of the first one. In this case,
%and based on our previously model
we focus on following the trajectories of the
bacteria´s fitness to model the evolutionary dynamics of the LTEE and
describe the fixation of beneficial mutations against drift processes.

\section{A model of DFE of new mutations in the LTEE}

To follow the evolution of the fitness of each of the 5 million trajectories
of bacteria present in the LTEE, the first reasonable parameters are the
probability per cell of occurring a mutation and the probability that it
be beneficial. Taking into account the values previously estimated for
fixed mutations in Ref.\cite{DarioyAugusto} and without differentiating between point
mutations and large rearrangements, an inferior cote for the parameters
would be: $p_{mut}=10^{-3}$ and $p_{b}=10^{-4}$, giving a net value for the probability of occurrence of a beneficial mutation of  $P_{b}=p_{mut}\times p_{b} =10^{-7}$. We took this numbers as a
starting point. In case of a beneficial mutation, a model described in Ref.\cite{Adaptation_Lenski} was used to increase the fitness, in which the advantage $s$ of the
mutation is distributed exponentially with a probability density of $\alpha
e^{-\alpha s}$. The advantage is defined from the values of fitness before
and after the mutation: $\omega^{\prime}=\omega (1+s)$. The $\alpha$ parameter also
change in time from a start value $\alpha_0$ as $\alpha^{\prime}=\alpha (1+gs)$ (only in case of beneficial
mutation). $g$ and $\alpha_0$ are fixed parameters for each of the twelve cultures of the
LTEE. They are interpreted as the initial DFE of the benefice, $\alpha_0$, and the epistasis parameter, $g$; worth $4.065$ and $60$ respectively for the population called Ara-1 \cite{Adaptation_Lenski}.

In case of mutation, but not a beneficial one, there are two possibilities:
neutral or deleterious. A new parameter is, in this case, the percent
that the non beneficial mutation is neutral, $p$, giving a rate of neutral mutations of: $P_{n}=p_{mut}\times{(1-p_{b})}\times p$. Here the fitness remains unchanged: $\omega^{\prime}=\omega$. The probability of deleterious mutation and its net rate, $P_d$ is
univocally determined by the others as $P_{d}=p_{mut}\times{(1-p_{b})}\times (1-p)$, and our proposal to
diminishing the fitness in this case is a linear one: $\omega^{\prime}=\omega r$,
where $r$ is a random real uniformly distributed in the interval [0.7, 1).
The inferior cote is arbitrary at first, but it represents a cutoff value from which
we considered depreciable the probability of a cell with a fitness so small
to survive trough seven generations (approximately one day of the
experiment). Firstly, we took the relation between this kind of mutations as the equality ($p=1/2$), later we check this is not true.

\begin{figure}
\includegraphics[width=8.5cm]{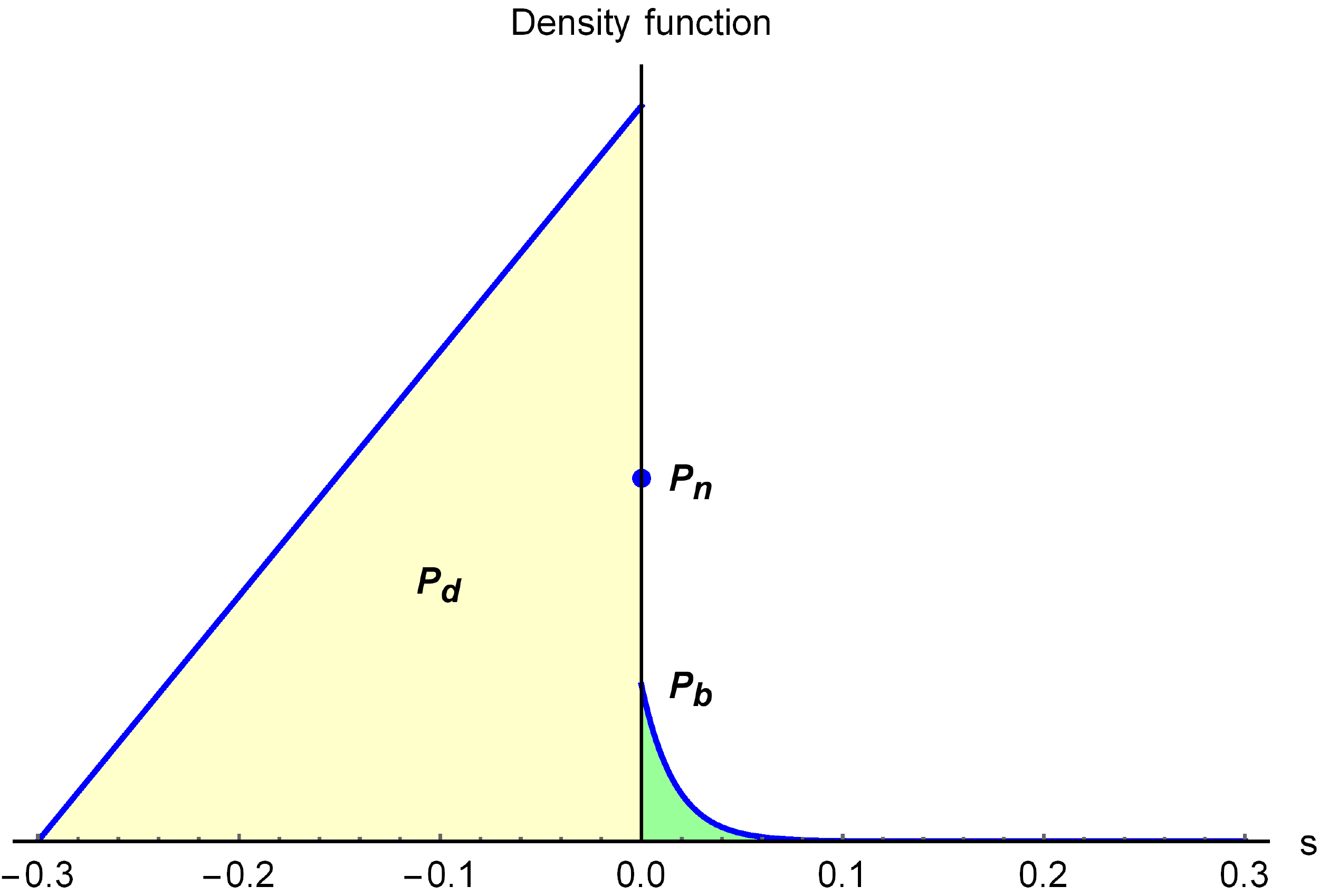}
\caption{A model of the density probability function of the fitness landscape in the LTEE depending of the net values of the probabilities of beneficial, neutral and deleterious mutation.}
\label{DFE}
\end{figure}

We have already constructed the DFE of all kind of mutations and
the density probability function can be viewed in Fig.\ref{DFE}.

\section{Dynamics of the LTEE}

\begin{figure}
\includegraphics[width=6.5cm]{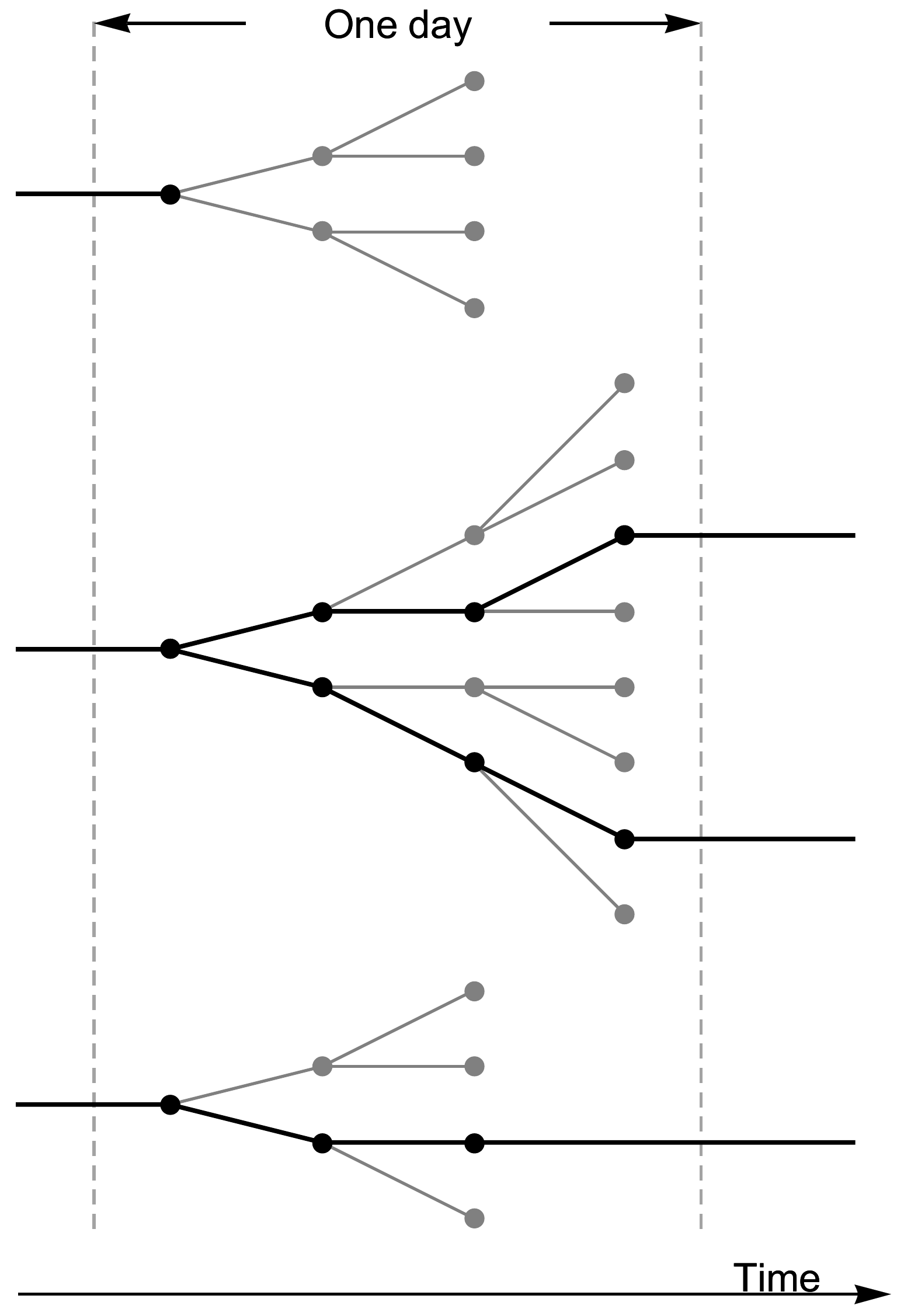}
\caption{Scheme of the one-day phylogeny of the LTEE. Only three of the 5 million lineages are represented by doing 2 or 3 replication steps of the 6 or 7 that they really are.}
\label{filo}
\end{figure}

The LTEE count with 12 different E. Coli populations with a common ancestor. Every day cultures of 5 million of bacteria are reproduced over 6 or
7 generations approximately rising up to 100 times the initial quantity. At
the end of the day, among the 500 million of resulting clones, a sample of $%
1\%$ is randomly selected to continue next day (see a better description in Ref.\cite{Sitio_Lenski}). The population size is regularly controlled. During almost 30 years a big
data on fitness has been recollected over 50 000 generation \cite{Sitio_Lenski}. The evolution
dynamics in the LTEE is schematically represented in Fig.\ref{filo}. Cell lineages
with deleterious mutations are usually truncated, whereas beneficial
mutations confer evolutionary advantage to clones and, thus, higher
probability to continue. Once they appear, some beneficial mutations are fixed in
more than $50\%$ of the population after a fixing time, others fail in the competence.

\begin{figure}
\includegraphics[width=7.0cm]{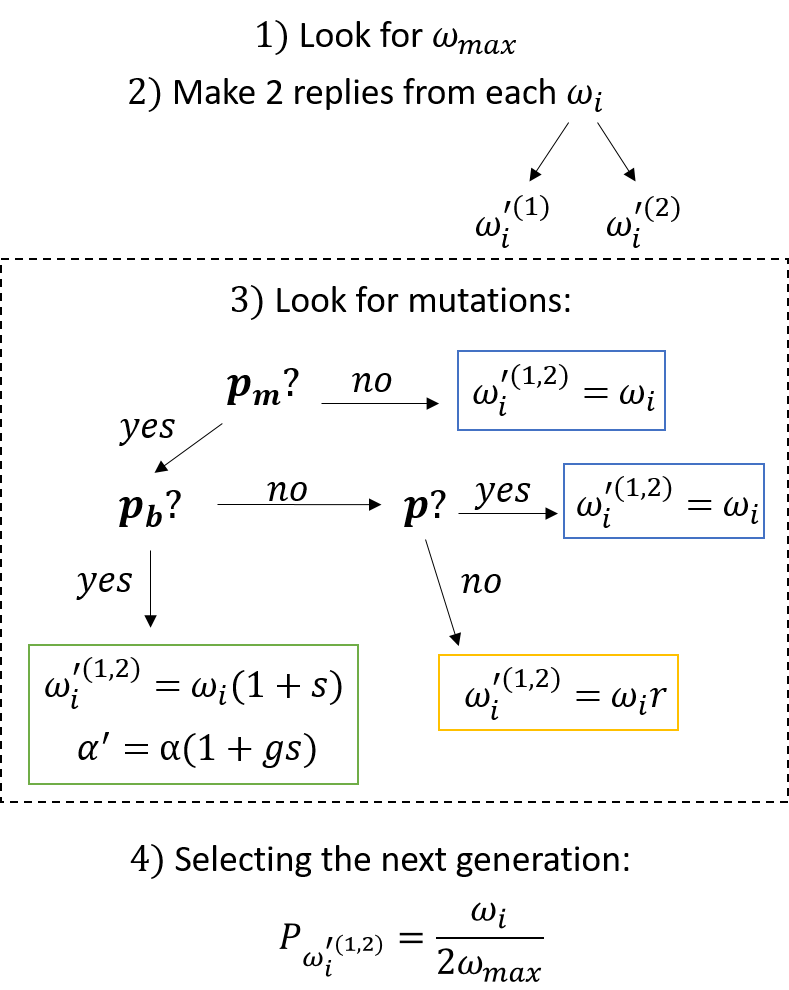}
\caption{Schematic pseudo-code of the algorithm run in the simulations. The advantage $s$ is distributed exponentially with a probability density of $\alpha e^{-\alpha s}$. Whereas the $r$ variable is uniformly distributed in the interval $[0.7, 1)$. Also, notice that the selection of each clone son, $\omega_i^{\prime (1,2)}$, depend only on the fitness of his progenitor, $\omega_i$.}
\label{algo}
\end{figure}

Merely because of a computational cost issue, and due to the fact that the number of cell lineages that continue in the experiment remains constant, we propose simply to simulate the LTEE with only one step of replication and the selection of half the clones. This mean by asserting that each linage produces exactly two clones in the replication step, but each son clone has a different probability of be selected, depending on his fitness, of course. So, that way we can maintain the advantage of clones with higher fitness to reach next generation. To perform this task  we design an algorithm that effectively reproduces the dynamics of the fitness in the experiment. Our proposal is the following one: first, we take the maximum fitness, $\omega_{max}$, as a reference. The second step is to make two copies from each cell, and to determine if there is any
mutation with his respectively probabilities. Then, we modify the fitness of the clones according to the model of DFE in Fig.\ref{DFE}. Next, the continuance or not of the clones sons of the ones with the maximum fitness is randomly, with
probability $1/2$. For the rest, the probability is weighted as: $\omega/2\omega_{max}$. A pseudo-code of the algorithm can be viewed in Fig.\ref{algo}.

The most of the times, after finishing this procedure have not been selected the 5 million of clones yet, a few thousands are missing. They are selected randomly from the total clones, because the relative frequencies of such a small number do not carry significant weight.

\begin{figure}
\includegraphics[width=8.5cm]{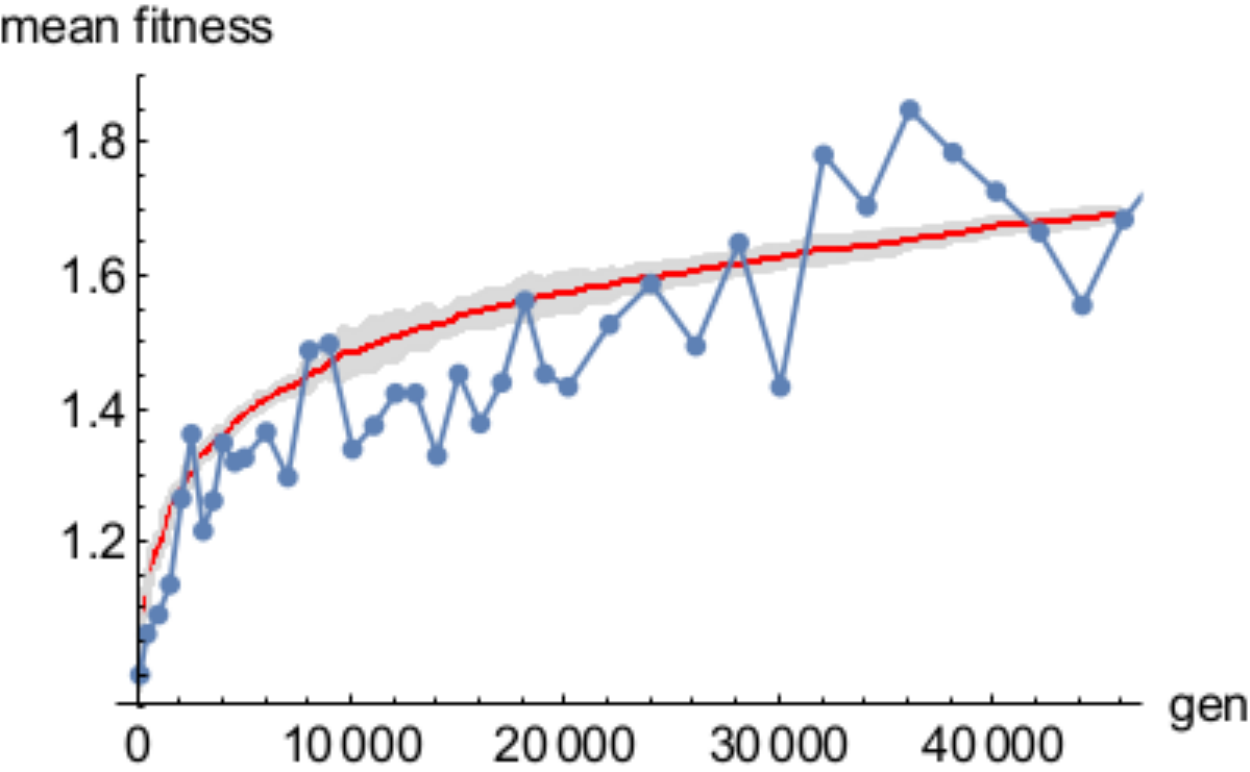}
\includegraphics[width=8.5cm]{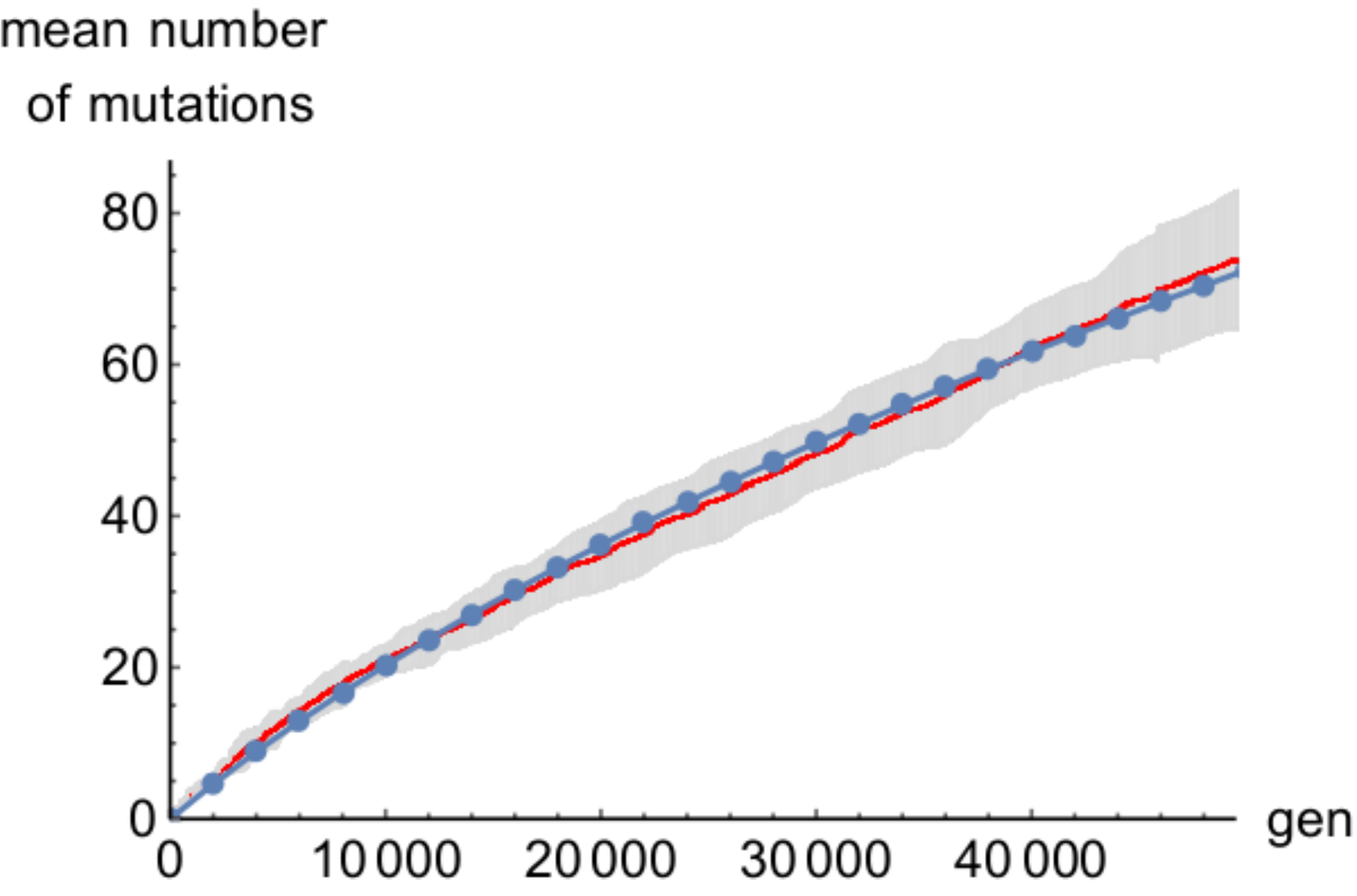}
\includegraphics[width=8.5cm]{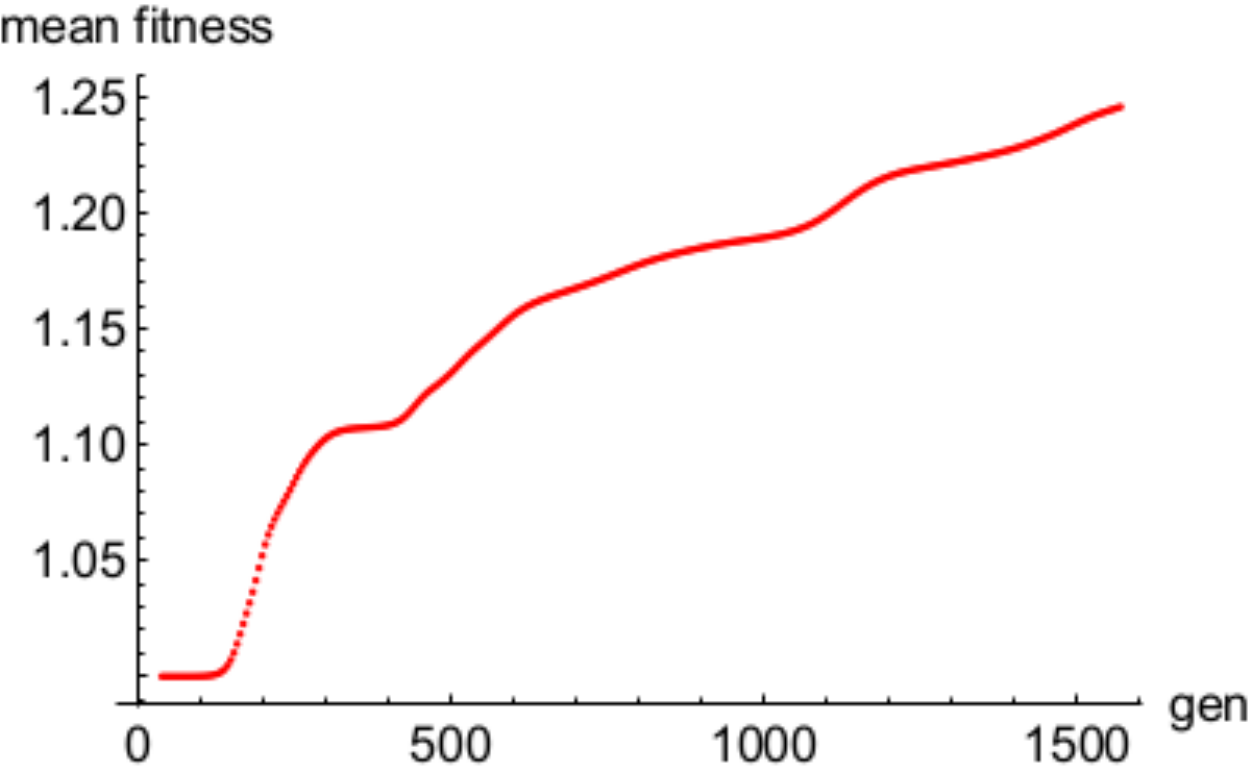}
\caption{a) Mean fitness along 50 000 generations of evolution in the LTEE. The  blue  point  corresponds  to  the experimental  data  and  the  red  ones  to  a mean of  six  of our simulations.  The  gray  zone  represents  the  deviation  of  the  simulations. b) Mean number of mutations along 50 000 generations of evolution in the LTEE. The  blue  point  corresponds  to an inferior cote computed from the experimental data and the red ones to  the mean of  six  of our simulations.  The  gray  zone  represents  the  deviation  of  the simulations. c) Amplified view of the mean fitness along the first 1 500 generations.}
\label{f-g}
\end{figure}

In the actual experiment the situation we have is that shown in Fig.\ref{filo}: for one day each lineage makes a different number of divisions, depending on the advantage of its fitness. The phenotype of the cells with the greatest advantage will grow faster and at the end of the day will be a little bit  more represented. That is why, although all cells are chosen with the same probability, the phenotypes better represented are more likely to be selected and pass the next day. The idea of the operation of our algorithm, restricted to one division, is that it represents the advantage of the phenotypes with high fitness weighting the selection for its fitness, in this way it guarantees the supremacy of the best adapted, as well as in the experiment. Our algorithm is just a reinterpretation of what the fitness concept means (see Ref.\cite{Fitness}).

As a resume, our algorithm has 4 free parameters: two of them to differentiate the types of mutations, $p_b$ and $p$, and another two referred to the model of beneficial mutations, $\alpha_0$ and $g$.
To test our model, we focus on the Ara-1 population and started the simulations with the values mentioned above. We then look for the optimal values of the parameters that best fit the experimental data on mean fitness and the total number of mutations at the same time. We had previously estimated an inferior cote for the mean number of mutations as function of the generation, $n$, which take the form: $4/3 \times (\sqrt{1+0.87\times 10^{-3} n}-1)/0.87$ for Ara-1 population (see Ref.\cite{DarioyAugusto}). We included the factor $4/3$ to account for both kinds of events: point mutations and large rearrangements. Fig.\ref{f-g}a y Fig.\ref{f-g}b respectively show the behavior of the mean fitness and the mean number of mutations in 50 000 generations coming from an average of six simulations with the optimal values, the experimental data fit very well. The set of values found is as follows: $p_b=8.2 \times 10^{-4}$, $p=0.8$, $\alpha_0=60$ and $g=6.0$. As a result of the simulation the value of $p_b$ is higher than the initial one, a possible explanation is the very low fixation rate of beneficial mutations, that mean there are much more beneficial mutations occurring in the experiment than the ones  we can measure (the fixed ones). Concerning non-beneficial mutations, neutrals are the $80\%$ and deleterious are the other $20\%$. It is known that approximately half of non-synonymous mutations are neutrals \cite{A2}. Synonymous mutations are neutral too by definition, because of the degeneration of the genetic code. Also, we can compute the quantity of the non-synonymous as the fraction of the $20$ amino acids that DNA codes of the 64 possible combinations of 3 of the 4 DNA nucleotides. With this data in mind, if we compute the quantity of neutral mutations, synonymous and non-synonymous, we will get a value of $84\%$. A rigorously measure is presented in \cite{rate}, which is in perfect agreement with the optimum value of the simulation. On the other hand, $\alpha_0$ kept his value, while $g$ rise up, but to a reasonable value too: the optimum $g$ math with the mean value of the 12 populations of the LTEE. The more sensitive parameter by far is $p_b$, in second place is $p$.

\section{Phenomenology}

\begin{figure}
\includegraphics[width=8.5cm]{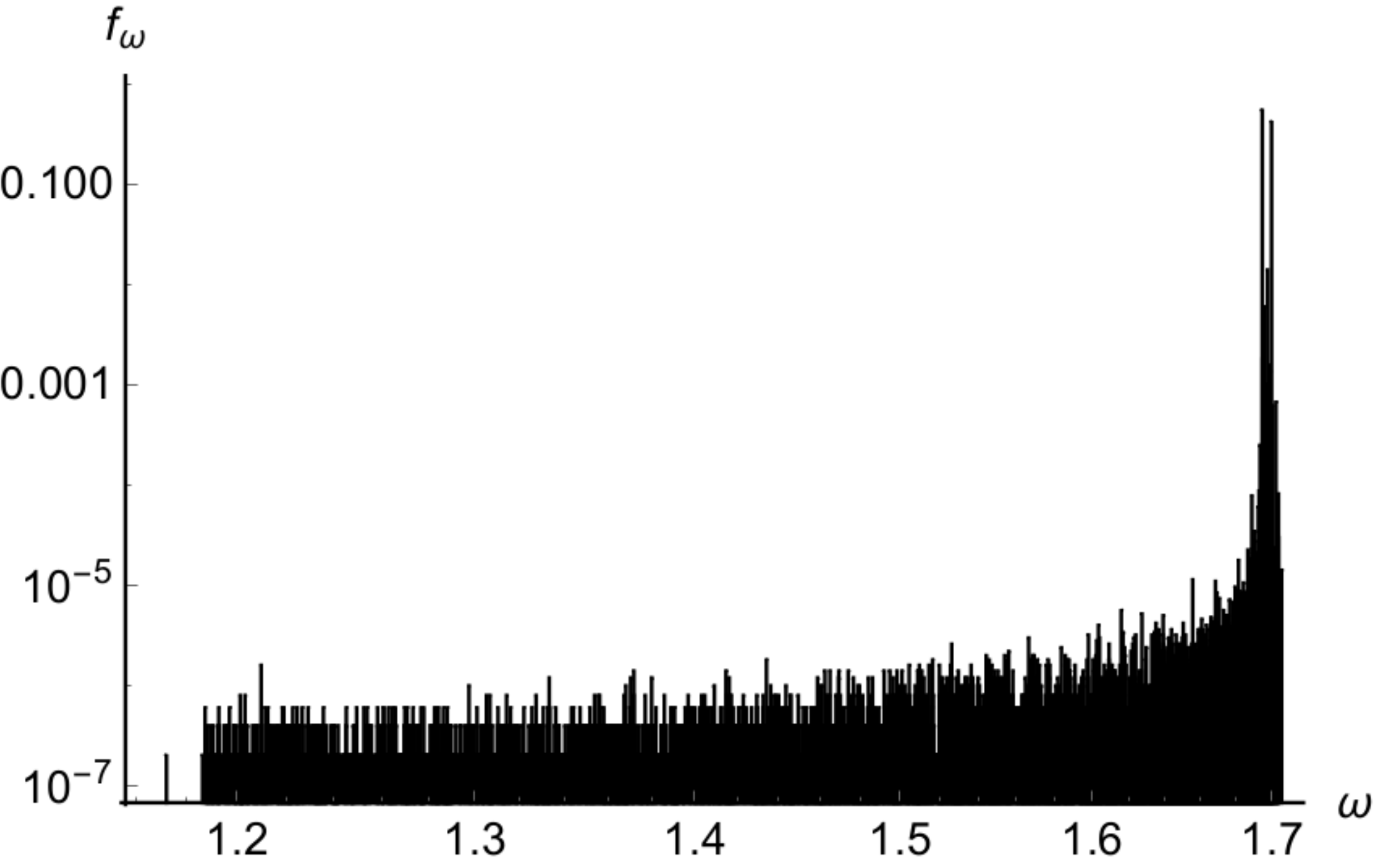}
\caption{Profile of the phenotypic variability in the generation 50 000 of a run of the LTEE.}
\label{perfil}
\end{figure}

If we take a closer view on the simulated curves we appreciate a lot of steps, as the mean fitness shows in Fig.\ref{f-g}c. This kind of behavior was suggested in Ref.\cite{20mil_Lenski}, but the size of the data uncertainty do not allow them to assure that. Each  step come from a fixed beneficial mutation and the  larger  the  step  the shorter the time of fixation: a larger advantage is more easily fixed. Also we can appreciate the sublinearity as a result of the epistasis phenomenon, which consist in the difficulty, more and more marked, of fixing a beneficial mutation on another previously fixed. Another consequence of epistasis is the reduction of the fitness´s deviation in the simulations. It turns out that for long times the mean fitness approaches its asymptotic behavior, being less likely to observe the deviation produced by the randomness of the simulation. On the contrary, the mean number of mutations keep growing and increasing its deviation. Another interesting issue is the clonal interference, which result of the competence of the clones to be the dominant one. Of course, not all impose themselves; the losers are source of phenotypic variability and there could be a lot of diversity (see Ref.\cite{A1}). In Fig.\ref{perfil} we can see the final profile of a simulation. There are two vast majority phenotypes, although there are others with higher fitness who star to compete. There could be moments with just one major phenotype, like in the instant after a fixation of a very beneficial mutation or in instants after the very beginning, where almost all cells have the same initial phenotype.

\section{Concluding remarks}

Despite there are multiple variants, the presented model of DFE of the mutations in Fig.\ref{DFE} is similar to the more accepted configuration in the literature (see Ref.\cite{DFE}), it has only one maximum located in the neutral point. Also it allows, by means of our algorithm, to simulate effectively the dynamic evolution of the LTEE. This can be used to better understand the working of evolution. The simulations can show some data unknown for the experiment, like the fraction of fixed mutations and a possible description of the phenotypic variability. This could be improved by massive sequencing the DNA of the clones.

\section{Acknowledgments}

 The authors acknowledge support from the National Program of Basic Sciences in
Cuba, and from the Office of External Activities of the
International Center for Theoretical Physics (ICTP).

%%%%%%%%%%%%%%%%%%%%%%%%%%%%%%%%%%%%%%%%%%%%%%%%%%%%%%%%%%%%%%%%%%%

\end{document}